\documentclass[prb,twocolumn,showpacs,preprintnumbers,amsmath,amssymb]{revtex4}
\usepackage{graphicx}

\begin{document}

\title{Theoretical chemistry of $\alpha$-graphyne: functionalization,
symmetry breaking, and generation of Dirac-fermion mass}

\author{R.~Longuinhos$^1$, E.~A.~Moujaes$^2$, S.~S.~Alexandre$^{1}$, and R. W. Nunes$^1$}
\affiliation{$^1$Departamento de F\'\i sica, ICEx, Universidade
Federal de Minas Gerais, 31270-901, Belo Horizonte, MG, Brazil\\
$^2$Departamento de F\'{i}sica, Universidade Federal de
Rond\^{o}nia, 76900-900, Porto Velho, Brazil}
\date{\today}

\begin{abstract}
We investigate the electronic structure and lattice stability of
pristine and functionalized (with either hydrogen or oxygen)
$\alpha$-graphyne systems. We identify lattice instabilities due to
soft-phonon modes, and describe two mechanisms leading to gap opening
in the Dirac-fermion electronic spectrum of these systems: symmetry
breaking, connected with the lattice instabilities, and partial
incorporation of an $sp^3$-hybrid character in the covalent-bonding
network of a buckled hydrogenated $\alpha$-graphyne lattice that
retains the symmetries of the parent pristine $\alpha$-graphyne. In
the case of an oxygen-functionalized $\alpha$-graphyne structure, each
O atom binds asymmetrically to two twofold-coordinated C atoms,
breaking inversion and mirror symmetries, and leading to the opening
of a sizeable gap of 0.22~eV at the Dirac point. Generally, mirror
symmetries are found to suffice for the occurrence of gapless Dirac
cones in these $\alpha$-graphyne systems, even in the absence of
inversion symmetry centers. Moreover, we analyze the gapless and
gapped Dirac cones of pristine and functionalized $\alpha$-graphynes
from the perspective of the dispersion relations for massless and
massive free Dirac fermions. We find that mirror-symmetry breaking
mimics a Dirac-fermion mass-generation mechanism in the
oxygen-functionalized $\alpha$-graphyne, leading to gap opening and to
isotropic electronic dispersions with a rather small electron-hole
asymmetry. In the hydrogen-functionalized case, we find that carriers
show a remarkable anisotropy, behaving as massless fermions along the
{\bf M}-{\bf K} line in the Brillouin zone and as massive fermions
along the $\boldsymbol\Gamma$-{\bf K} line.
\end{abstract}

\maketitle

\section{Introduction}
Monolayer and few-layer graphene have been under intensive scientific
scrutiny since their isolation and identification on a variety of
substrates~\cite{Novoselov22102004,Sutter2008406,Li20091312} This
research effort has also generated an increasing interest in the
identification of other two-dimensional (2D) materials that could
possibly share the linear dispersion and chiral nature of the
electronic carriers in graphene, along with the ensuing exceptional
electronic properties.~\cite{rmp,NovoselovGeim2005,jenaina}
Topological insulators,~\cite{tpli} silicene,~\cite{silicene}
monolayer boron-nitride,~\cite{bn} and monolayers of inorganic layered
materials such as molybdenum sulfide~\cite{mos2}, have since come into
the fold.
\begin{figure}
\includegraphics[width=8.5cm]{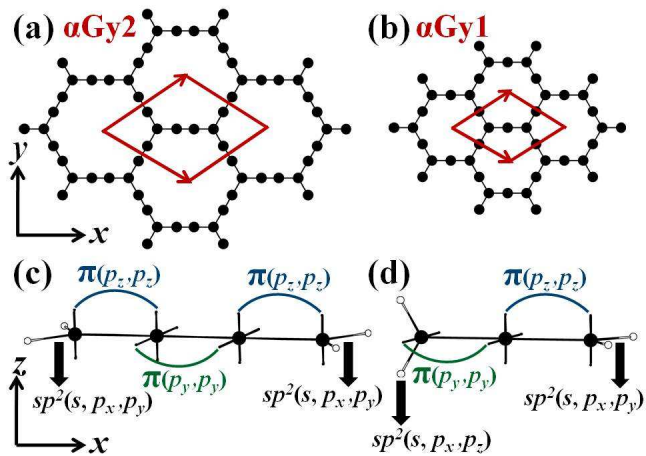}
\caption{{\large $\alpha$}-graphyne ({\large $\alpha$}Gy) 2D lattices
and corresponding linear molecular forms.  (a) {\large $\alpha$}Gy2
with two twofold carbon atoms in the linear chains connecting the
threefold sites. Lattice vectors and the eight-atom primitive cell
are indicated. (b) {\large $\alpha$}Gy1 with one twofold carbon atom
in the linear chains. Lattice vectors and the five-atom primitive
cell are indicated. (c) Butratiene: C$_4$H$_4$ hydrocarbon with
co-planar methylene (CH$_2$) groups at the edges. (d) Allene:
C$_3$H$_4$ hydrocarbon with perpendicular methylene groups at the
edges. In (c) and (d), the $\pi$-bonding scheme along the chains and
the $sp^2$ bonding of the methylene groups are indicated.}
\label{fig1}
\end{figure}

Nevertheless, the all-carbon ``sector'' of this problem has not been
fully exhausted, and families of two-dimensional (2D) carbon
allotropes named graphynes have also been investigated for the
occurrence of Dirac cones of electronic
states.~\cite{grphy,PhysRevLett.108.086804} While synthesis of
crystalline forms of graphynes is still lacking, many chemical routes
have been reported for the large molecules that form the building
blocks for such crystals~\cite{nat-review,prod,ADMA:ADMA200902623}. A
few recent works have examined these graphyne systems, related
hydrocarbons, and other functionalized
forms.~\cite{PhysRevB.86.045443,C3CP53237B} Further in-depth
theoretical investigation of the electronic properties and lattice
stability of such carbon allotropes is a crucial step in gauging the
potential applications of graphynes as graphene-like materials.

Among a variety of possible graphyne
lattices~\cite{PhysRevLett.108.086804,grphy,PSSB:PSSB201046583}, the
so-called $\alpha$-graphyne ({\large $\alpha$}Gy) shares with graphene
the hexagonal Bravais lattice and the presence of two
threefold-coordinated carbon atoms that occupy the sites of a 2D
honeycomb lattice. In {\large $\alpha$}Gy, the threefold carbons are
connected by linear chains of two twofold-coordinated carbon atoms, as
shown in Fig.~\ref{fig1}(a). Linear chains with a different number of
twofold carbon atoms connecting the threefold honeycomb sites are also
conceivable. Figure~\ref{fig1}(b) shows the case of one-atom chains.

In any {\large $\alpha$}Gy lattice, for any length of the
twofold-carbon chains, an inversion-symmetry center in the middle of
any given linear chain relates the twofold-coordinated atoms
two-by-two (except for one carbon atom that sits at the inversion
center in odd-membered chains) and the two threefold sites are
inversion-symmetry partners as well. From this perspective, inversion
symmetry could be believed to be a requirement for the occurrence of
Dirac cones in the electronic dispersion of $\alpha$-graphynes, as in
graphene. This question was touched upon in
Ref.~\onlinecite{PhysRevLett.108.086804}, where it was shown by explicit
calculations for the so-called 6,6,12-graphyne that a lattice need not
belong to the $p6m$ hexagonal-symmetry group to display Dirac cones in
its electronic dispersion. The 6,6,12-graphyne lattice belongs to the
$pmm$ rectangular symmetry group, yet Dirac cones appear in its
electronic structure, albeit not at the {\bf K}-point in the Brillouin
zone (BZ).~\cite{PhysRevLett.108.086804} Given that both of these
lattice-symmetry groups display twofold-symmetry axes (which are
equivalent to inversion centers in 2D lattices), the question of
whether inversion symmetry is a requirement for the occurrence of
Dirac cones still stands.

In this work, we examine the nature of the electronic states of planar
$\alpha$-graphynes with linear chains of one [{\large $\alpha$}Gy1 in
Fig.~\ref{fig1}(b)] and two [{\large $\alpha$}Gy2 in
Fig.~\ref{fig1}(a)] twofold carbon atoms, and discuss the
chemical-bond picture that determines the chemical stabilization of
these carbon allotropes by the saturation of in-plane $p$ orbitals,
promoted by covalent binding of either hydrogen or oxygen with the
twofold carbons along the chains, yielding: (i) an {\large
$\alpha$}Gy2-based hydrocarbon showing Dirac cones of electronic
states at the Fermi energy ($E_f$); and (ii) an {\large
$\alpha$}Gy2-based graphyne oxide that meets the oft-quoted need of
the presence of a gap in the electronic dispersion at the Fermi level.

Furthermore, we investigate the lattice stability of graphyne
hydrocarbons based on both {\large $\alpha$}Gy1 and {\large
$\alpha$}Gy2, aiming at identifying possible lattice instabilities of
the 2D geometries and their impact on the electronic dispersions of
these systems. We also examine symmetry requirements for the presence
of Dirac cones of massless fermions in the electronic spectrum of
these systems, as well as the connection between symmetry breaking and
the presence of gaps, or lack thereof, in the Dirac-cone spectrum of
pristine and functionalized {\large $\alpha$}Gy1 and {\large
$\alpha$}Gy2. We describe two different mechanisms for gap opening in
the Dirac cones of $\alpha$-graphyne systems: (i) symmetry breaking;
(ii) incorporation of an $sp^3$-hybrid content in the covalent-bonding
network of the twofold C atoms.

We find that while a pristine 2D {\large $\alpha$}Gy2 lattice is fully
stable, with no soft modes in the phonon spectrum, a planar form of an
{\large $\alpha$}Gy2+H hydrocarbon, with C$_8$H$_6$ stoichiometry,
shows a soft-phonon mode at the $\boldsymbol\Gamma$ point in the
phonon BZ that drives the system to a fully stable (no soft-phonon
modes) buckled geometry, with off-plane displacements of the twofold
carbons and the saturating hydrogen atoms, as shown in
Fig.~\ref{fig2}(b). The soft-mode displacements preserve the inversion
centers and mirror symmetries of the planar {\large $\alpha$}Gy2+H
lattice, yet we observe the opening of a gap of $\sim$0.05~eV in the
Dirac cone at $E_f$, which is related to the partial incorporation of
an $sp^3$ character in the covalent bonds of the twofold C atoms.  The
gapped Dirac cone at $E_f$ in the buckled {\large $\alpha$}Gy2+H
lattice is strongly anisotropic, with a linear dispersion (hence
massless fermions) along the {\bf M-K} direction in the BZ and a
quadratic dispersion (hence massive fermions) along the 
{\bf K-}$\boldsymbol\Gamma$ direction.

In the case of the planar {\large $\alpha$}Gy2+O system with
C$_8$O$_3$ stoichiometry, each O atom binds asymmetrically to the two
twofold C atoms in the chains, as shown in Fig.~\ref{fig2}(c),
breaking both inversion and mirror symmetries and leading to the
opening of a sizeable gap of 0.22~eV in the Dirac cone at $E_f$.

In the case of {\large $\alpha$}Gy1, the pristine planar {\large
  $\alpha$}Gy1 lattice is found to be chemically unstable, due to the
formation of in-plane $\pi$ bonds that leads to frustration of the
planar geometry. Saturation of in plane $p$ orbitals, by hydrogen
functionalization, inhibits the formation of the destabilizing $\pi$
bonds, and produces a planar {\large $\alpha$}Gy1+H geometry
(C$_5$H$_3$ stoichiometry) that lacks inversion symmetry but keeps the
mirror planes cutting the lattice plane through the CH bonds. The
mirror planes are found to suffice for the occurrence of gapless Dirac
cones in the electronic structure of this 2D {\large $\alpha$}Gy1+H
geometry. Moreover, this planar geometry shows a lattice instability,
due to a soft-phonon mode at the $\boldsymbol\Gamma$ point, leading to
a buckled structure with off-plane displacements of twofold carbons
and hydrogen atoms, as shown in Fig.~\ref{fig2}(a). Mirror-symmetry
breaking, connected with the soft-mode displacements, leads to gap
openings ranging from $0.16$ to $0.60$~eV at the Dirac points in the
electronic spectrum of this system.

Furthermore, we analyze gapless and gapped Dirac cones of the 
{\large $\alpha$}Gy2, {\large $\alpha$}Gy2+H, and {\large $\alpha$}Gy2+O
systems, from the perspective of the dispersion relations for massless
and massive free Dirac fermions. We find that the electronic bands of
the gapped {\large $\alpha$}Gy2+O system are isotropic, display a
small ($\sim$5-6\%) electron-hole asymmetry, and are very well
described by the relativistic energy-momentum dispersion relation for
free massive Dirac fermions, suggesting that symmetry breaking
operates as a mechanism of mass generation for the Dirac charge
carriers in {\large $\alpha$}Gy2+O. In the case of the buckled {\large
$\alpha$}Gy2+H system, we find a strongly anisotropic behavior, with
carriers acting as massless fermions along the {\bf M-K} line in the
BZ and as massive fermions along the $\boldsymbol\Gamma${\bf -K} line.
Moreover, effective hopping values in $\alpha$-graphynes are computed
by employing the standard relations between hopping integrals, {\bf
M}-point van Hove-singularity gaps, and Fermi velocities obtained from
the standard tight-binding description of graphene, in order to
explain the reduction of carrier velocities in $\alpha$-graphynes,
when compared with graphene.

\section{Methods}
The electronic structure analysis and structural optimization of the
{\large $\alpha$}Gy systems were carried out using the SIESTA code
\cite{soler2002siesta}. Geometry relaxation was performed until the
total force on each atom was lower than 4$\times 10^{-2}$ eV/\AA, and
pressures were lower than 0.5 Kbar. We used a 64$\times$64$\times$1
Monkhorst-Pack (MP)~\cite{PhysRevB.13.5188} {\bf k}-point sampling of
the BZ, and an atomic basis set with two polarization zeta functions
(DZP), with an energy shift of 0.01~Ry and a mesh cutoff of 250~Ry.
Trouiller-Martins norm-conserving pseudopotentials in the fully
non-local scheme were employed for the interaction between valence
electrons and ionic cores~\cite{hamann1979norm,PhysRevLett.48.1425}.
The SIESTA calculations were performed using a van der Waals
functional (vdW-DF)~\cite{roman2009efficient,PhysRevLett.92.246401}
and double checked with the PBE GGA
functional~\cite{perdew1996generalized}. The two functionals lead to
similar results, as expected for these systems.

The study of lattice stability was performed within the density
functional perturbation theory~\cite{RevModPhys.73.515}, as
implemented in the Quantum Espresso (QE)
code~\cite{0953-8984-21-39-395502}. In these calculations we further
relaxed the Siesta-relaxed structures until the forces were lower than
2$\times 10^{-3}$ eV/\AA\ and the pressure was lower than 0.5 Kbar.  A
plane-wave energy cutoff of 32~Ry was found to produce well-converged
results. Ionic cores were represented by ultrasoft
pseudopotentials~\cite{PhysRevB.41.1227,PhysRevB.41.7892}. We used the
PBE GGA functional~\cite{perdew1996generalized} and double checked the
calculations for negative frequencies using the LDA functional. The
Fermi surface was smeared using cold
smearing~\cite{PhysRevLett.82.3296} with a degauss of 0.01.  We used a
{\bf k}-point sampling grid of 12$\times$12$\times$1 for the
electronic states and a {\bf q}-point sampling grid of
4$\times$4$\times$1 for the phonon calculations. The threshold for the
calculations of phonon spectra was 10$^{-21}$.

\section{Results and discussion}
\subsection{Chemical-bond constraints on the stability of 
planar $\alpha$-graphynes}
We start by outlining chemical considerations indicating that pristine
$\alpha$-graphynes with odd numbers of twofold carbon atoms along the
chains should be unstable in the planar form, due to chemical
frustration. This can be understood by considering the $sp$-hybrid
covalent bonding of the following molecular hydrocarbons known as
cumulenes: butatriene (C$_4$H$_4$) and hexapentaene (C$_6$H$_4$), that
are stable in a planar form, and allene (C$_3$H$_4$) and pentatetraene
(C$_5$H$_4$), with an extended-tetrahedral structure where the
methylene groups at the two terminations of the molecule are
perpendicular to each
other~\cite{doi:10.1021/ja00252a002,doi:10.1021/j100302a014}. For the
sake of the argument, it suffices to consider butatriene and allene,
shown in Figs.~\ref{fig1}(c) and (d), respectively.
\vspace{0.5cm}
\begin{figure} [htb]
\includegraphics[width=8.5cm]{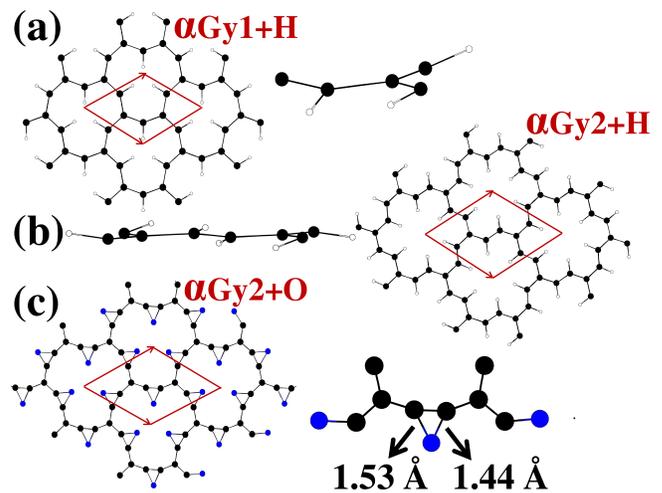}\\
\caption{Functionalized $\alpha$-graphyne structures. (a) Left side:
top view of planar {\large $\alpha$}Gy1+H; lattice vectors and the
primitive cell with C$_5$H$_3$ stoichiometry are indicated. Right
side: lateral view of the C$_5$H$_3$ unit of the buckled geometry,
showing asymmetric off-plane displacements of C and H atoms. (b) Right
side: top view of planar {\large $\alpha$}Gy2+H; lattice vectors and
the primitive cell with C$_8$H$_6$ stoichiometry are indicated. Left
side: lateral view of the C$_8$H$_6$ unit of the buckled geometry,
showing symmetric off-plane displacements of C and H atoms. (c) Left
side: top view of planar {\large $\alpha$}Gy2+O; lattice vectors and
the primitive cell with C$_8$O$_3$ stoichiometry are indicated. Right
side: planar C$_8$O$_3$ unit showing asymmetric bonding of the O atom
with the two C atoms in the linear chain. C, O, and H atoms are shown
as black, blue, and white circles, respectively.}
\label{fig2}
\end{figure}

In Figs.~\ref{fig1}(c) and (d), the axis of each molecule lies along
the $x$ direction, and the twofold atoms in the chain form $\sigma$
bonds along the chain axis, involving $sp_x$ hybrids. In the case of
allene in Fig.~\ref{fig1}(d), with a single twofold carbon atom
connecting the threefold carbons at each end of the molecule, the
plane formed by the methylene group (CH$_2$) on the left end must be
perpendicular to that on the right end of the molecule. This follows
from the fact that whenever the covalent double bond between the
twofold C atom and the threefold one on the right involves a $\pi$
bond between $p_z$ orbitals, the double bond with the threefold atom
on the left must necessarily involve a $\pi$ bond between the $p_y$
orbitals of the two atoms. This means that when CH bonds on the right
are ($s$,$p_x$,$p_y$) $sp^2$ hybrids, those on the left must be
($s$,$p_x$,$p_z$) $sp^2$ hybrids, as illustrated in
Fig.~\ref{fig1}(d), leading to the chemical frustration of the planar
form. This is true for any odd number of atoms in the chain connecting
the two threefold carbons. For even-membered chains, a similar
argument shows that the methylene groups at the two ends lie on the
same plane, as shown in Fig.~\ref{fig1}(c). This translates into
chemical-bond constraints on the stability of extended geometries:
stable planar forms are not expected for odd-membered chains.

The {\large $\alpha$}Gy1 and {\large $\alpha$}Gy2 lattices are the
extended analogs of allene and butatriene, respectively.  We performed
an {\it ab initio} structural optimization of the {\large $\alpha$}Gy1
system, starting from a fully planar geometry, that resulted in a
highly distorted structure, due to chemical frustration, while {\large
$\alpha$}Gy2 retained a 2D form and the honeycomb lattice upon
structural relaxation, providing confirmation of the above
chemical-bond picture. Throughout this report, we address the lattice
stability of the hydrogen-functionalized {\large $\alpha$}Gy systems
we propose. The lattice stability of pristine {\large $\alpha$}Gy2 has
been addressed in Ref~\cite{doi:10.1021/jp3111869}, with no unstable
phonon modes found in this system, a result that we have
confirmed in our calculations.

The bond chemistry outlined in Figs~1(c) and (d) also suggests that
in-plane functionalization of $\alpha$-graphynes should lead to
stabilization of planar forms, at least from the perspective of the
chemical saturation of the in-plane $p$ orbitals of the twofold
carbons in the chains. Below, we consider in-plane functionalization
with either hydrogen or oxygen, and address the electronic structure
and lattice stability of the planar forms.

\subsection{Electronic structure of pristine $\alpha$-graphyne}
\begin{figure} [htb]
\includegraphics[width=8.5cm]{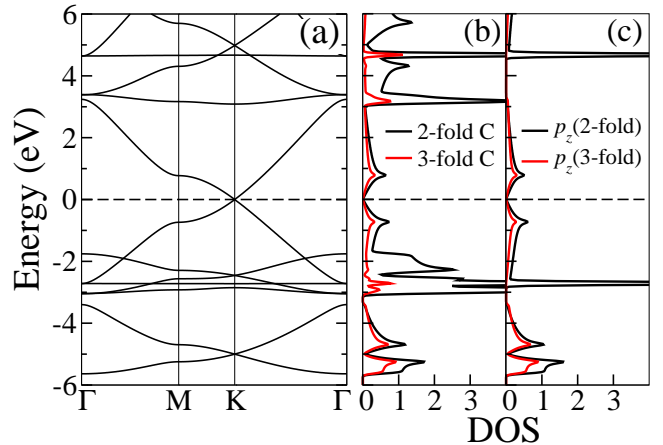}\\
\caption{(a) Band structure of pristine {\large $\alpha$}Gy2 along the 
$\boldsymbol\Gamma${\bf -M}, {\bf M-K}, and 
{\bf M-}$\boldsymbol\Gamma$ high-symmetry lines in the Brillouin zone. 
(b) Black line: partial density of states (PDOS) projected on
all basis orbitals of the six twofold C atoms. Red line: PDOS
projected on all orbitals of the two twofold C atoms. (c) Black
line: PDOS projected on the $p_z$ orbitals of the six twofold C
atoms. Red line: PDOS projected on the $p_z$ orbitals of the two
twofold C atoms.}
\label{fig3}
\end{figure}

Let us first consider the nature of the electronic bands and the
corresponding density of states (DOS) of the pristine {\large 
$\alpha$}Gy2 structure, as shown in Fig.~\ref{fig3}. In the figure,
we also show the partial density of states (PDOS) for the twofold and
threefold C atoms. We consider {\large $\alpha$}Gy2 monolayers on the
$xy$ plane as in Figs.~\ref{fig1}(a) and (b). The remarkable
similarity of the {\large $\alpha$}Gy2 electronic structure with that
of graphene itself has been pointed out in recent
works~\cite{PhysRevLett.108.086804,PhysRevB.86.115435,doi:10.1021/jp4067795}. 
Here, we take a closer look at the {\large $\alpha$}Gy2 electronic
dispersion and discuss where it differs from that of graphene.
\vspace{0.8cm}

In graphene, the two $p_z$ orbitals per unit cell give rise to the two
$\pi$ bands crossing each other at $E_f$. In {\large $\alpha$}Gy2,
there are eight C atoms per unit cell, and we expect eight bands
connected with the eight $p_z$ orbitals in the cell.  Regarding the
Dirac cone at $E_f$ in Fig.~\ref{fig3}, altogether the orbitals of the
six twofold atoms account for $\sim$65-70\% of the related DOS, and
the two threefold C atoms contribute the remaining $\sim$30-35\%,
showing that the $p_z$ orbitals of the twofold C atoms resonate with
the $p_z$ orbitals of the threefold C atoms to form the metallic $\pi$
band. This clearly indicates that the presence of an inversion
symmetry center in the middle of the twofold-carbon chain and an
underlying hexagonal-symmetry lattice enable the appearance of Dirac
cones of resonating $\pi$ orbitals even for the more complex atomic
basis of {\large $\alpha$}Gy2, when compared with graphene.

Figure~\ref{fig3} also shows additional Dirac cones with the same
overall band topology as the one at $E_f$, one with its Dirac point at
$\sim$5.0~eV below $E_f$, with nearly the same orbital composition,
and another at $\sim$5.0~eV above $E_f$, the latter being derived from
linear combinations of the in-plane ($p_x$ and $p_y$) orbitals which
are perpendicular to the twofold-carbon chains. Moreover, strongly
localized molecular-like bonding and antibonding combinations of these
in-plane $p$ orbitals of the twofold C atoms are also observed, as
indicated by the nearly dispersionless bands that appear as sharp
Dirac-delta-like peaks in the DOS, at $\sim$3.3~eV above and
$\sim$3.0~eV below $E_f$. These localized states have no $p_z$ orbital
character, as shown by the PDOS curves in Figs.~\ref{fig3}(b) and
(c). Localized states derived from the $p_z$ orbitals also appear as
two flat bands one at $\sim$2.7~eV below and another at $\sim$4.6~eV
above $E_f$ [Figs.~\ref{fig3}(b) and (c)].

\subsection{Hydrogen functionalization of $\alpha$-graphyne: 
electronic structure and lattice stability}
The above chemical-bond and electronic-structure analysis suggests
possible ways of adding other atomic species such as hydrogen or
oxygen in order to form stable {\large $\alpha$}Gy2 compounds. The
in-plane $p$ orbitals are natural candidates to form CH bonds with H
atoms added in the plane, in the voids of the {\large $\alpha$}Gy2
lattice. The localized character of the molecular-like states derived
from these in-plane orbitals suggests a strong reactivity to
functional groups. In the case of {\large $\alpha$}Gy1, bond chemistry
suggests that the chemical frustration of the planar structure
observed in the pristine {\large $\alpha$}Gy1 lattice should be
suppressed by passivating the in-plane $p$ orbitals, thus inhibiting
the formation of the destabilizing $\pi$ bonds from these orbitals.
\\
\vspace{0.5cm}
\begin{figure} [htb]
\includegraphics[width=8.5cm]{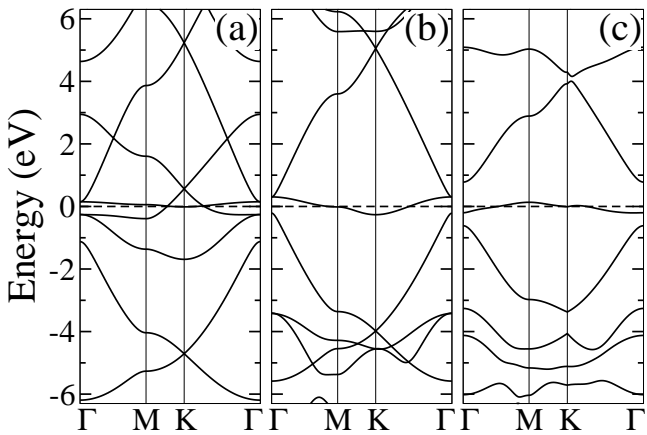}\\
\caption{(a) Band structure of a constrained (see text) planar
pristine {\large $\alpha$}Gy1 lattice along high-symmetry lines in the
Brillouin zone. (b) Band structure of a planar functionalized {\large
$\alpha$}Gy1+H lattice. (c) Band structure of a buckled functionalized
{\large $\alpha$}Gy1+H lattice showing gap openings on the Dirac cones
at the {\bf K}-point.}
\label{fig4}
\end{figure}

Indeed, {\it ab initio} structural relaxation of a {\large $\alpha$}Gy1+H 
hydrocarbon (C$_5$H$_3$ stoichiometry), where in-plane H atoms bond to
the twofold C atoms in the chains, leads to the planar structure shown
in Fig.~\ref{fig2}(a). To be precise, in our calculations the planar
{\large $\alpha$}Gy1+H is ``numerically stable'', in the sense that
starting from an initial planar geometry where all C and H atoms lie
on the $xy$ plane, the structure relaxes into a planar geometry that
retains the honeycomb-like lattice and shows Dirac cones in its
electronic structure, as displayed in Fig.~\ref{fig4}(b). Note,
however, that no Dirac cone is found at $E_f$, where we observe only a
band with a small dispersion, connected with localized states derived
from the $p_z$ orbitals of the twofold C atoms. This nearly
dispersionless band at $E_f$ is likely a feature of the electronic
structure of any odd-membered {\large $\alpha$}Gy lattice.

For comparison, Fig.~\ref{fig4}(a) shows the electronic bands for a
pristine {\large $\alpha$}Gy1 geometry (without H atoms) that was
constrained to remain in the planar hexagonal lattice, where we also
observe the band of localized states crossing the Fermi level, and the
occurrence of Dirac cones at the {\bf K}-point, in energies above and
below $E_f$. In this case, we observe a cone with its Dirac point at
$\sim$0.5~eV above $E_f$ and a dispersive band just below $E_f$, both
composed nearly entirely of the $p_x$ and $p_y$ orbitals of the
twofold C atoms. Upon hydrogen saturation of the in-plane $p$
orbitals, these bands are shifted away from the Fermi level, as shown
in Fig.~\ref{fig4}(b).

The picture drawn above of the chemical stability of the planar
{\large $\alpha$}Gy1+H provides only a heuristic argument, and does
not exclude the possibility of lattice instabilities due to
soft-phonon modes in the phonon dispersion relation for this
structure. In order to address this issue, we computed the phonon
frequencies at the $\boldsymbol\Gamma$ point in the BZ for the planar
{\large $\alpha$}Gy1+H and found two soft modes with frequencies of
$\omega = -377,74$~cm$^{-1}$ and $\omega =-376.51$~cm$^{-1}$.  By
displacing the atoms according to the lowest-frequency soft mode, the
system relaxes onto the structure shown on the right in
Fig.~\ref{fig2}(a), where H and C atoms shift off-plane, but the
structure retains an underlying hexagonal lattice, with all threefold
C atoms placed at the sites of a honeycomb lattice. The off-plane
shifts of the twofold carbons and H atoms are asymmetrical, with one C
(H) atom shifting ``above'' the plane by $\sim$0.21\AA\ (0.71\AA), and
the other two C (H) atoms shifting ``below'' the plane by
$\sim$0.21\AA\ (0.69\AA) and $\sim$0.16\AA\ (0.52\AA),
respectively. Hence, in this geometry both inversion centers and
mirror planes are broken.
\\
\vspace{0.1cm}
\begin{figure} [htb]
\includegraphics[width=8.5cm]{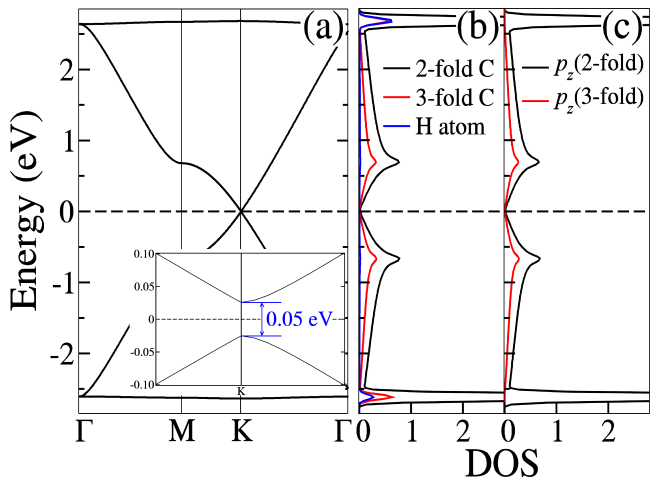}\\
\caption{(a) Band structure of functionalized planar {\large
$\alpha$}Gy2+H along high-symmetry lines in the Brillouin zone.  The
inset shows gap opening ($\sim$0.05 eV) on the Dirac point at the
Fermi energy, that occurs in the buckled geometry. (b) Black line:
partial density of states (PDOS) projected on all basis orbitals of
the six twofold C atoms. Red line: PDOS projected on all orbitals of
the two threefold C atoms. Blue line: PDOS projected on all orbitals
of the six H atoms (c) Black line: PDOS projected on the $p_z$
orbitals of the six twofold C atoms. Red line: PDOS projected on the
$p_z$ orbitals of the two threefold C atoms.}
\label{fig5}
\end{figure}

The band structure for this buckled {\large $\alpha$}Gy1+H is shown in
Fig.~\ref{fig4}(c), where we observe gap openings at the {\bf K}-point 
ranging from 0.16~eV for the Dirac cone above $E_f$ to 0.6~eV for the
one below $E_f$. Gap opening in this case is connected with the
breaking of the mirror planes which are present in the planar
form. The band structure of {\large $\alpha$}Gy1+H suggests a poor
electronic material with small carrier mobilities associated with the
nearly dispersionless band straddling the Fermi level. While stable
with respect to $\boldsymbol \Gamma$-point phonon modes, the buckled
{\large $\alpha$}Gy1+H structure is itself unstable against soft modes
in other phonon $\vec{q}$-vectors, such as the {\bf K} and {\bf M}
points in the BZ, but we do not pursue these lattice instabilities in
this report.
\begin{table}[htbp]
\centering
\begin{tabular}{|l|c|c|c|}
\hline System &$\boldsymbol\Gamma$ &{\bf K} &{\bf M}\\ \hline 
{\large $\alpha$}Gy1+H (planar) &-378 , -377 &\textendash &\textendash \\ \hline 
{\large $\alpha$}Gy1+H (buckled) &stable &-159 &-103 \\ \hline 
{\large$\alpha$}Gy2 (planar) &stable &stable &stable \\ 
            &(full) &(full) &(full) \\ \hline 
{\large $\alpha$}Gy2+H (planar) &-104 &stable &stable \\ \hline 
{\large $\alpha$}Gy2+H (buckled) &stable &stable &stable \\ 
            &(full) &(full) &(full) \\ \hline 
\end{tabular}
\caption{Frequencies (in cm$^{-1}$) of unstable phonon modes computed
in special {\bf k}-points in the Brillouin zone, in pristine and
H-functionalized $\alpha$-graphyne lattices. ``Stable'' indicates
absence of unstable phonon modes at the {\bf k}-point indicated.  For
pristine {\large $\alpha$}Gy2 and the buckled {\large $\alpha$}Gy2+H
structures full dispersions along the $\boldsymbol\Gamma${\bf -K},
{\bf K-M}, and $\boldsymbol\Gamma${\bf -M} high-symmetry lines in the
Brillouin zone were computed (indicated by ``full'' in parenthesis).}
\label{phonon}
\end{table}

We turn now to the more interesting case of the related {\large
$\alpha$}Gy2+H hydrocarbon, with C$_{\rm 8}$H$_{\rm 6}$
stoichiometry. In the {\large $\alpha$}Gy2 lattice, by bonding one H
atom to each twofold C atom, a planar structure that retains the
inversion symmetry center in the middle of the chain is formed, as
shown on the right in Fig.~\ref{fig2}(b). The electronic bands for
this planar {\large $\alpha$}Gy2 hydrocarbon, shown for a narrower
energy interval in Fig.~\ref{fig5}(a), display the characteristic
Dirac cone at $E_f$ as in pristine {\large $\alpha$}Gy2. The main
differences between the band structures of pristine {\large
$\alpha$}Gy2 and {\large $\alpha$}Gy2+H in this interval are connected
with the saturation of the in-plane $p$ orbitals which move the
non-dispersive bands derived from these orbitals away from the Fermi
level, making for a ``cleaner'' electronic structure in an interval of
$\pm$5.0~eV centered on $E_f$. In this range, we observe only a cone
with a Dirac point at $E_f$, shouldered by a pair of
molecular-orbital-like non-dispersive bands, derived mostly from the
$p_z$ orbitals of the twofold C atoms, at $\pm$~2.6 eV from $E_f$.

Regarding lattice stability, the planar {\large $\alpha$}Gy2+H is
unstable against a soft-mode phonon at the $\boldsymbol \Gamma$ point
(we found no unstable modes at the {\bf K} and {\bf M} points for this
structure), and geometry relaxation starting from an initial geometry
where atomic positions are displaced according with the soft-phonon
mode leads to the geometry shown on the left in Fig.~\ref{fig2}(b),
where H and twofold C atoms shift out of the plane, resulting in a
buckled geometry where some degree of $sp^3$ bonding is incorporated
in the C-C and C-H bonds of the twofold carbons. As a result, we
observe the opening of a 0.05~eV gap in the Dirac cone, as shown in
the inset in Fig.~\ref{fig5}(a).
\vspace{0.5cm}
\begin{figure} [htb]
\includegraphics[width=8.5cm]{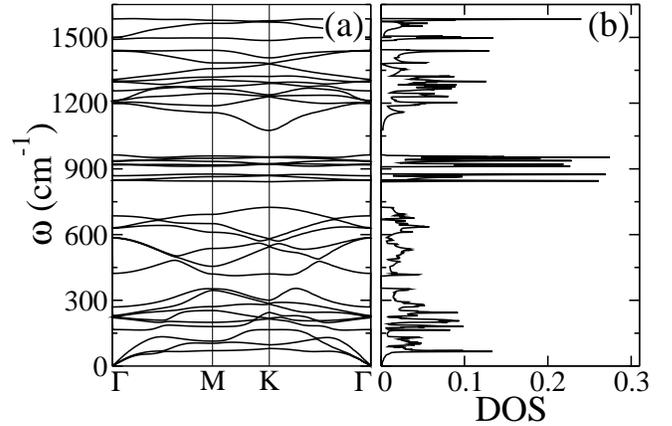}\\
\caption{(a) Phonon spectrum of the buckled {\large $\alpha$}Gy2+H
lattice. Six additional phonon branches, spanning the frequency
range of 3082-3108 cm$^{-1}$, are not shown. (b) Phonon density of
states for the buckled {\large $\alpha$}Gy2+H lattice.}
\label{fig6}
\end{figure}

We observe that in this case the system retains the hexagonal lattice,
the inversion centers, and the mirror planes after the out-of-plane
relaxation. As in the buckled {\large $\alpha$}Gy1+H, the threefold C
atoms occupy the sites of a planar honeycomb lattice. Note that the
gap at $E_f$ in this buckled {\large $\alpha$}Gy2+H structure, due to
the mixed hybridization state (mostly $sp^2$ with some $sp^3$
character on the twofold-carbon bonds), is one order of magnitude
smaller than the gap in the buckled {\large $\alpha$}Gy1+H and in the
planar {\large $\alpha$}Gy2+O systems (the latter is discussed below).

The full lattice stability of this buckled {\large $\alpha$}Gy2+H
geometry is confirmed by the calculation of the phonon dispersion and
phonon DOS shown in Fig.~\ref{fig6}, where no negative frequencies are
observed. In the figure, the phonon dispersion is shown along the
$\boldsymbol\Gamma${\bf -K}, {\bf K-M}, $\boldsymbol\Gamma${\bf -M}
high-symmetry lines in the BZ. In Table~\ref{phonon} we summarize the
results of our calculations of phonon frequencies in {\large
$\alpha$}Gy systems.

\subsection{Oxygen functionalization of $\alpha$-graphyne: 
electronic structure, symmetry breaking, and gap opening} 
We switch now to the functionalization of {\large $\alpha$}Gy2 with O
atoms. Oxygen is known to bind to the graphene sheet in two
configurations, depending on the local coverage~\cite{nanot}. The only
stable form at large oxygen coverages is the unzipped configuration,
where the O atom is located above the graphene layer, and binds with
two C atoms by breaking their mutual C-C bond. Our chemical-bond
analysis indicates that a stable oxygenated {\large $\alpha$}G2 form
may be expected where the O atoms bind with the twofold carbons
through the in-plane $p$ orbitals as in the hydrogenated cases
above. In this case, however, one O atom is enough to chemically
saturate the pair of twofold C atoms in each chain.  We consider thus
an {\large $\alpha$}Gy2 oxide with C$_{\rm 8}$O$_{\rm 3}$
stoichiometry~\cite{C3CP53237B}.
\\
\begin{figure} [htb]
\includegraphics[width=8.5cm]{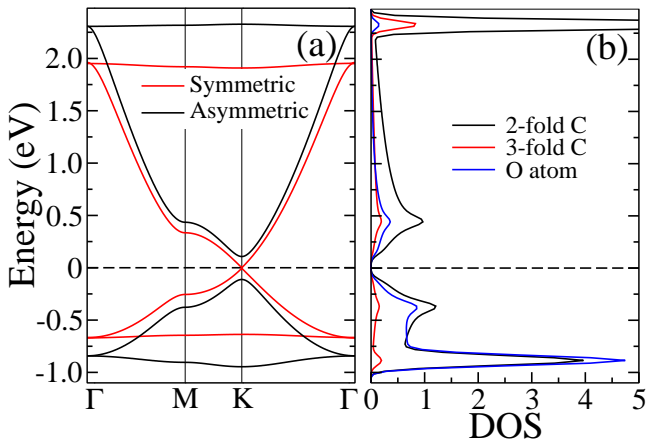}\\
\caption{(a) Band structure of the functionalized planar {\large
$\alpha$}Gy2+O structure along high-symmetry lines in the Brillouin
zone. Black lines: electronic bands for the fully relaxed asymmetric
geometry, showing gap opening (0.22 eV) at the Fermi energy. Red
lines: electronic bands for a mirror-symmetric geometry, showing no
gap opening at the Fermi energy. (b) DOS for the asymmetric
structure. Black line: partial density of states (PDOS) projected on
all basis orbitals of the six twofold C atoms. Red line: PDOS
projected on all orbitals of the two threefold C atoms. Blue line:
PDOS projected on all orbitals of the three O atoms.}
\label{fig7}
\end{figure}

The resulting planar structure is shown in Fig.~\ref{fig3}(c). The
cooperative strain mechanism that leads to the unzipped phase of
oxygen on graphene~\cite{PhysRevLett.96.176101} in not operative in
this case, and the O atom makes bonds with the two carbons in the
so-called clamped
configuration~\cite{nanot,PhysRevLett.103.086802}. Unlike in the case
of {\large $\alpha$}G2+H, both inversion and mirror symmetries are
broken in this case, as shown on the right in Fig.~\ref{fig2}, leading
to gap opening in the Dirac cone at $E_f$, as seen in
Fig.~\ref{fig7}(a). Note that the valence and conduction bands are
highly dispersive, indicating low effective masses and hence the
possibility of high carrier mobilities in this material. This feature,
together with the sizeable gap of 0.22~eV, makes this an interesting
candidate for a material with a graphene-like gapped dispersion. A
full account of the lattice stability of this {\large $\alpha$}G2+O
will be the subject of a forthcoming study.

\section{Symmetry considerations and gap opening}
Symmetry considerations are in order at this point. The pristine
{\large $\alpha$}Gy1 lattice, when numerically constrained to a planar
honeycomb-like lattice, shows twofold axes at the positions of the
twofold carbons (i.e., 2D inversion centers) and mirror planes, hence
belongs to the 2D $p6m$ hexagonal-symmetry group.  In the relaxed
planar geometry of the functionalized {\large $\alpha$}Gy1+H, the
threefold carbons sit on the sites of a perfectly planar honeycomb
lattice. However, inversion symmetry is broken in this structure,
while the mirror-symmetry planes (intersecting the sheet through the
CH bonds) are retained, as can be seen in Fig.~\ref{fig2}(a). Thus,
{\large $\alpha$}Gy1+H belongs to the $p31m$ hexagonal-symmetry group, yet we
observe the occurrence of Dirac cones at the {\bf K}-point in the
Brillouin zone, as shown in Fig.~\ref{fig4}(b).

While inversion symmetry is usually regarded as essential for the
occurrence of Dirac cones of electronic states (both the
hexagonal-lattice {\large $\alpha$}Gy2 and the rectangular-lattice
{\large $\beta$}-graphyne considered in
Ref.~\onlinecite{PhysRevLett.108.086804} display inversion centers), even in
the absence of inversion symmetry, mirror symmetry ensures the
occurrence of Dirac cones in the electronic structure of the planar
{\large $\alpha$}Gy1+H system, as shown in
Fig.~\ref{fig4}(b). Mirror-symmetry breaking in the buckled {\large
  $\alpha$}Gy1+H lattice reduces the symmetry of this system to the
$p3$ hexagonal-symmetry group and leads to gap opening in the Dirac
cones at the {\bf K}-point.

Note that, from the point of view of the transformation of the $p_z$
orbitals under these symmetries, the mirror planes and inversion
centers differ only by a phase. In a related result, in
Ref.~\onlinecite{PhysRevB.81.073408} a mirror symmetry was shown to be
connected with the occurrence of Dirac cones on the electronic states
associated with grain boundaries in graphene.

In the case of {\large $\alpha$}Gy2, both the planar and buckled forms
of {\large $\alpha$}Gy2+H belong to the $p6m$ symmetry group and
gap opening in the buckled lattice is not connected with symmetry
breaking, as discussed above. 

The {\large $\alpha$}Gy2+O case is similar to {\large $\alpha$}Gy1+H:
a mirror-symmetry-constrained planar lattice displays mirror symmetry
planes and no inversion centers, thus belonging to the $p31m$ group,
and gapless Dirac cones appear in its electronic spectrum
[Fig.~\ref{fig7}(a)], while in the unconstrained lattice mirror
symmetry is broken, the overall symmetry is reduced to the $p3$ group,
and we observe gap openings at the Dirac points.

\subsection{Massive Dirac-fermion nature of carriers in 
functionalized $\alpha$-graphynes} 
Let us focus now on the Dirac-fermion nature of the electronic
carriers in these $\alpha$-graphyne systems. That the carriers in the
{\large $\alpha$}Gy2+O lattice can be described as massive Dirac
fermions is confirmed by the excellent fittings we obtain for the
dependence of the energy of the Dirac-like bands to the relativistic
dispersion relation for massive fermions:
\begin{equation}
E^2\left(\vec{k}\right) = \hbar^2 k^2 v_F^2\; + m^2
v_F^4\;;
\label{mass-fit}
\end{equation}
where $v_F$ and $m$ are, respectively, the effective ``speed of
light'' and mass of the Dirac-fermions. A natural choice is to set the
rest-energy term to $ m^2 v_F^4 = (E_g/2)^2$, where $E_g$ is the gap
that opens at the Dirac point. With that choice, by fitting the energy
bands of this system to Eq.~\ref{mass-fit}, we obtain the carrier
velocities in Table~\ref{table2}. Included in the table are the
electron and holes velocities along the {\bf M-K} and 
{\bf K}-$\boldsymbol\Gamma$ directions in the BZ. We remark that $v_F$
values in Table~\ref{table2} show that the bands of 
{\large $\alpha$}Gy2+O are essentially isotropic with a rather small
($\sim$5-6\%) electron-hole asymmetry.

The opposite is true in the case of the buckled {\large $\alpha$}Gy2+H
lattice. The dispersion relation near the Fermi energy is strongly
anisotropic, displaying a linear behavior along the {\bf M-K}
direction and a quadratic one along the {\bf K}-$\boldsymbol\Gamma$
direction, as can be seen in the inset in Fig.~\ref{fig5}. Hence,
carriers in this system should behave as massless Dirac fermions along
the {\bf M-K} line and as massive fermions along the 
{\bf K}-$\boldsymbol\Gamma$ line.

Also included in Table~\ref{table2} are the Fermi velocities for the
massless Dirac fermions of pristine {\large $\alpha$}Gy2 and the
planar {\large $\alpha$}Gy2+H geometry, obtained from the fitting of
the corresponding gapless dispersion relations to
$E\left(\vec{k}\right) = \hbar\;\! k\;\!v_F$.
\begin{table}[htbp]
\centering
\begin{tabular}{|l|c|c|c|c|c|c|c|}\hline 
&\multicolumn{2}{c}{$v^e_F$} \vline
&\multicolumn{2}{c}{$v^h_F$} \vline 
&\multicolumn{2}{c}{$m$} \vline 
&$t_{eff}$ \\ \hline
System &{\bf M-K} &{\bf K}-$\boldsymbol\Gamma$ &{\bf M-K} &{\bf K}-$\boldsymbol\Gamma$ 
&{\bf M-K} &{\bf K}-$\boldsymbol\Gamma$ & \\ \hline 
{\large $\alpha$}Gy2 &0.671 &0.696 &0.667 &0.692  &0 &0 &0.75 \\ \hline 
{\large $\alpha$}Gy2+H (planar) &0.568 &0.598 &0.568 &0.595 &0 &0 &0.68 \\ \hline 
{\large $\alpha$}Gy2+H (buckled) &0.568 &0.611 &0.561 &0.611 &0 &0.01 &0.71 \\ \hline 
{\large $\alpha$}Gy2+O (planar) &0.349 &0.349 &0.329 &0.333 &0.18 &0.18 &0.41 \\ \hline 
\end{tabular}
\caption{Carrier velocities $v^e_F$ (electrons) and $v^h_F$ (holes),
in units of 10$^6$~m/s, along the {\bf M-K} and 
{\bf K}-$\boldsymbol\Gamma$ directions in the Brillouin zone, and values of
carrier mass $m$, in units of bare electron mass ($m_{bare}$), in
pristine and functionalized {\large $\alpha$}Gy2 lattices, from
fittings to relativistic dispersion relations for massless and massive
Dirac fermions. Effective hopping $t_{eff}$ (in eV), from
tight-binding model.}
\label{table2}
\end{table}

From Table~\ref{table2}, we observe that a sizeable renormalization of
the Fermi velocity takes place in the pristine {\large $\alpha$}Gy2
lattice ($v_F = 0.68\times 10^6$~m/s), when compared with graphene
($v_F = 1.00 \times 10^6$~m/s). In a simple orthogonal first-neighbor
tight-binding (TB) picture, the Fermi velocity of the Dirac fermions
is given by $v_F = \sqrt{3}\;\!t\;\!a_0/2\;\!\hbar$, where $t$ is the
hopping parameter and $a_0$ is the lattice constant.

An effective hopping parameter for each of these systems can be
defined from the gap $\Delta_M$ between the two van Hove singularities
at the {\bf M}-point in the BZ, i.e., $t_{eff} = \Delta_M/2$, as given
by this simple TB model. The values of the $t_{eff}$ in each case,
obtained from this prescription, are also included in
Table~\ref{table2}. We observe that the renormalization of $v_F$ in
these {\large $\alpha$}Gy2 lattices is connected with a strong
reduction of the effective hopping parameter, from $\sim$3~eV in
graphene to 0.41-0.75~eV in $\alpha$-graphynes.

\section{Conclusions}
In conclusion, through {\it ab initio} calculations we investigate the
electronic structure and the lattice stability of pristine and
functionalized (with either H or O atoms) $\alpha$-graphyne systems
with one ({\large $\alpha$}Gy1) and two ({\large $\alpha$}Gy2) atoms
along the twofold-carbon chains. We identify their lattice
instabilities, connected with soft-phonon modes, and describe two
mechanisms leading to gap opening in the Dirac-fermion electronic
spectrum of these systems: symmetry-breaking connected with the
lattice instabilities and partial incorporation of an $sp^3$ character
in the bonding network of the hydrogenated {\large $\alpha$}Gy2+H,
where the fully-lattice-stable buckled geometry retains the symmetries
of the parent pristine {\large $\alpha$}Gy2.

More specifically, our calculations indicate that a 2D lattice of
pristine {\large $\alpha$}Gy2 is fully stable, displaying no soft
phonon modes, while its planar hydrogenated counterpart {\large 
$\alpha$}Gy2+H is unstable against a $\boldsymbol\Gamma$-point
soft-phonon mode. The instability drives the system to a fully-stable
buckled geometry, where twofold C atoms and saturating H atoms shift
off the plane of the threefold C atoms. In the case of {\large 
$\alpha$}Gy1, while a pristine 2D lattice is found to be unstable
due to chemical-bond frustration, a planar {\large $\alpha$}Gy1+H
geometry is chemically stabilized by saturation of in plane $p$
orbitals that inhibits the formation of the destabilizing $\pi$
bonds. The planar {\large $\alpha$}Gy1+H lacks inversion symmetry but
retains mirror planes that are found to suffice for the occurrence of
gapless Dirac cones in the electronic dispersion of this system.
Mirror-symmetry breaking and concurrent gap openings in the Dirac
cones take place in a buckled {\large $\alpha$}Gy1+H lattice.

In the case of a planar {\large $\alpha$}Gy2+O system, a single O atom
bind asymmetrically to the two C atoms in the twofold chain, breaking
inversion and mirror symmetries, and leading to the opening of a
sizeable gap of $\sim$0.22~eV in the Dirac cone at the Fermi level.

Finally, in this study we also analyze the gapless and gapped Dirac
cones of the {\large $\alpha$}Gy2, {\large $\alpha$}Gy2+H, and {\large
$\alpha$}Gy2+O systems from the perspective of the dispersion
relations for massless and massive free Dirac fermions.We find that
mirror-symmetry breaking operates as a Dirac-fermion mass-generation
mechanism in {\large $\alpha$}Gy2+O, leading to gap opening and to
isotropic electronic dispersions with a rather small electron-hole
asymmetry. In {\large $\alpha$}Gy2+H, carriers display a remarkable
anisotropy, acting as massless fermions along the {\bf M}-{\bf K} line
in the Brillouin zone and as massive fermions along the
$\boldsymbol\Gamma$-{\bf K} line.  Renormalization of carrier
velocities {\large $\alpha$}Gy2 lattices is found to be due to a
strong reduction of the effective hopping parameter, when compared
with graphene.\\

{\bf Acknowledgement}\\
We acknowledge funding from Brazilian agencies CNPq, FAPEMIG, Capes,
and Instituto Nacional de Ci\^{e}ncia e Tecnologia (INCT) em
Nanomateriais de Carbono - MCT.\\

\bibliography{graphyne}
\end{document}